\documentclass[12pt]{iopart}
\usepackage{epsfig}
\newcommand{\ANP}{{\it Adv. Nucl. Phys.} }
\newcommand{\EPJ}{{\it Eur. Phys. J.} }
\newcommand{\JHEP}{{\it JHEP} }
\newcommand{\MPL}{{\it Mod. Phys. Lett.} }
\newcommand{\PPNP}{{\it Prog. Part. Nucl. Phys} }
\newcommand{\PRP}{{\it Phys. Rep.} }
\newcommand{\SJNP}{{\it Sov. J. Nucl. Phys.} }
\newcommand{\beq}{\begin{equation}}
\newcommand{\eeq}{\end{equation}}
\newcommand{\bea}{\begin{eqnarray}}
\newcommand{\eea}{\end{eqnarray}}
\newcommand{\bce}{\begin{center}}
\newcommand{\ece}{\end{center}}

\def\lsim{\mathrel{\rlap{\lower4pt\hbox{\hskip1pt$\sim$}}
    \raise1pt\hbox{$<$}}}         %less than or approx. symbol
\def\gsim{\mathrel{\rlap{\lower4pt\hbox{\hskip1pt$\sim$}}
    \raise1pt\hbox{$>$}}}         %greater than or approx. symbol

\begin{document}

\title{The Vector Probe in Heavy-Ion Reactions} 

\author{Ralf Rapp\footnote[3]{email: rapp@comp.tamu.edu}  
}

\address{Cyclotron Institute and Physics Department, Texas A\&M University, 
               College Station, Texas 77843-3366, U.S.A.}

\begin{abstract}
We review essential elements in using the $J^P=1^-$ channel as a probe
for hot and dense matter as produced in (ultra-) relativistic collisions
of heavy nuclei. The uniqueness of the vector channel resides in the 
fact that it directly couples to photons, both real and virtual (dileptons),
enabling the study of thermal radiation and in-medium 
effects on both light ($\rho, \omega, \phi$) and heavy ($\Psi, \Upsilon$) 
vector mesons.  We emphasize the importance of interrelations between
photons and dileptons, and characterize relevant energy/mass regimes
through connections to QGP emission and chiral symmetry restoration.
% as well as lattice QCD calculations.
Based on critical analysis of our current understanding of
data from fixed-target energies, we identify open key questions
to be addressed.
% theoretically, as well as with the help of new measurements at RHIC and SPS.       
\end{abstract}

%Uncomment for PACS numbers title message
%\pacs{00.00, 20.00, 42.10}

% Uncomment for Submitted to journal title message
%\submitto{\JPG}

% Comment out if separate title page not required
%\maketitle

%%%%%%%%%%%%%%%%%%%%%%%%%%%%%%%%%%%%%%%%%%%%%%%%%%%%%%%%%%%%%%%%%%%%%%%%
\section{Introduction: Towards QGP Discovery}
\label{sec_intro}
%%%%%%%%%%%%%%%%%%%%%%%%%%%%%%%%%%%%%%%%%%%%%%%%%%%%%%%%%%%%%%%%%%%%%%%%
Collisions of heavy nuclei at high energies provide a rich laboratory 
for studying strongly interacting matter under extreme conditions
of (baryon-) density ($\varrho_B$) and temperature ($T$). One of the main 
objectives is the creation and identification of new forms of matter,
most notably Quark-Gluon Plasma (QGP), 
as predicted by the underlying theory, Quantum Chromodynamics (QCD). 

As a first step towards this goal, a necessary
condition for the investigation of the phase diagram of QCD 
is an approximate (local) thermalization of the produced 
matter.  Bulk properties of hadron production over a wide range of 
collision energies have indeed revealed ample evidence for  
multiple reinteractions, justifying the notion of the production of 
``matter". Among the main features that distinguish heavy-ion ($A$-$A$) 
collisions from elementary proton-proton ($p$-$p$) reactions are a 
strong collective expansion as extracted from hadronic transverse
momentum ($p_t$) spectra~\cite{HS98,KH03,Bass98,Brat00,Zhang00} and 
(apparent) chemical equilibration inferred from produced hadron 
species~\cite{Beca01,BRS03}, most notably in the strangeness sector.      
Furthermore, all measurements of dilepton invariant-mass spectra 
thus far~\cite{dls97,ceres160,helios3,na50,ceres40} 
exhibit large excess production in central $A$-$A$ collisions over 
$p$-$p$ (or $p$-$A$) reference spectra. 
In fact, at energies available at the Super Proton Synchrotron (SPS), 
theoretical analyses imply~\cite{RW00} that 
at least 10 generations of $\pi^+\pi^-\to e^+e^-$ annihilation are 
required to reproduce the observed dilepton yields in semi-central 
$Pb$(158AGeV)-$Au$ collisions. 

In a second step, one needs to assess thermodynamic state variables 
(temperature, energy density, pressure) characterizing the 
produced system. Already with $Pb$ beams at top SPS energy 
there have been several indications~\cite{HJ00} for temperatures 
($T\ge 200$~MeV) and energy densities ($\epsilon\ge 3$~GeV/fm$^3$)
significantly above the critical values obtained
from current lattice simulations of finite-$T$ QCD~\cite{KL03}.  
Recent data from the Relativistic Heavy-Ion Collider (RHIC) at about 
a factor 10 higher (center-of-mass) collision energy have added 
spectacular new results on bulk matter properties: 
(i) the suppression of high-$p_t$ particles 
(``jet-quenching")~\cite{GVWZ03}, requiring energy densities
$\epsilon\simeq30$~GeV/fm$^3$ in the early stages
of central $Au$-$Au$ collisions;
(ii) radial and especially elliptic flow of hadrons at low 
$p_t\le 2$~GeV (comprising more than 95\% of the total multiplicity)  
that are well-described by hydrodynamic simulations~\cite{KH03,TLS01}
favoring formation times of thermalized matter of 
$\tau_0\simeq 0.5$~fm/c implying initial temperatures 
$T_0\simeq 350$~MeV and energy densities quite consistent with (i);
(iii) an unexpectedly large baryon-to-meson ratio, as well as
an approximate ``constituent-quark scaling" of the elliptic flow of 
all measured hadrons, at intermediate $p_t\simeq3-6$~GeV, which
are both naturally explained by quark coalescence models
at the hadronization transition~\cite{Hwa03,Greco03,Fries03}.

The third step consists of understanding the nature of the created
matter, that is, its microscopic properties including  
phase changes to establish that a ``new" state has indeed been 
observed. In the present context, this means that one should identify 
properties of the QGP that go beyond, say, an ordinary electromagnetic
plasma. Examples of these include:
(1) the $L^2$ dependence of parton energy loss ($L$: path length of 
traversed matter of a high-energy parton), representing 
its non-abelian character;
(2) the origin of the partonic interactions that allow the system
to thermalize on the required short time scales (e.g.,
resonance states in the QGP as recently found in lattice 
QCD, or gluon multiplication processes~\cite{XG04} involving 
3- and 4-gluon vertices)  
(3) deconfinement;
(4) restoration of the spontaneous breaking of chiral symmetry (SBCS).
The last two items are basic features of the high-temperature QCD
phase transition that are also well established by lattice 
simulations~\cite{KL03}. To infer these from ultrarelativistic 
heavy-ion collisions (URHICs), the vector probe is expected to 
play a central role.
Its uniqueness resides on the fact that it carries the quantum
numbers of the photon, which does not undergo strong interactions
and thus can carry undistorted information on the hot and dense
phases of the fireball, both lightlike ($\gamma$) and
timelike (lepton pairs, $\gamma^*\to l^+l^-$
with $l=e,\mu$). Whereas photon emission is a suitable observable
to assess the temperature of the matter, dileptons encode additional
dynamical information through their invariant mass, $M$. In particular,
in the low-mass region they directly couple to the light vector mesons
and thus reflect their mass distribution at the moment
of decay, rendering them the prime observable to study mass (de-) 
generation related to (the restoration of) SBCS.
At higher mass, dileptons are a standard tool to measure the 
abundance of the heavy vector mesons ($\Psi$ and $\Upsilon$ families),
a systematic study of which is hoped to provide information on deconfinement,
see Ref.~\cite{RG04} for a recent overview.  

The article is organized as follows: in Sec.~\ref{sec_emrad} we recall
basic features of thermal electromagnetic (e.m.) emission in  heavy-ion 
collisions.  In Sec.~\ref{sec_dilep} we give a detailed discussion of 
the physics potential of thermal dileptons (including 
comparisons to data), both at low mass in connection
with chiral restoration and at intermediate mass in connection with
QGP radiation, and summarize their prospects for RHIC. 
In Sec.~\ref{sec_phot} we review the current status in the assessment
of photon production rates from hot and dense matter, and again
put the results into context with measurements at SPS and RHIC.
A summary including a list of open questions is given in 
Sec.~\ref{sec_concl}.

%%%%%%%%%%%%%%%%%%%%%%%%%%%%%%%%%%%%%%%%%%%%%%%%%%%%%%%%%%%%%%%%%%%%%%%%
\section{Four Pillars of Electromagnetic Radiation}
\label{sec_emrad}
\begin{figure}[!t]
\vspace{-1.5cm}
\begin{center}
\epsfig{file=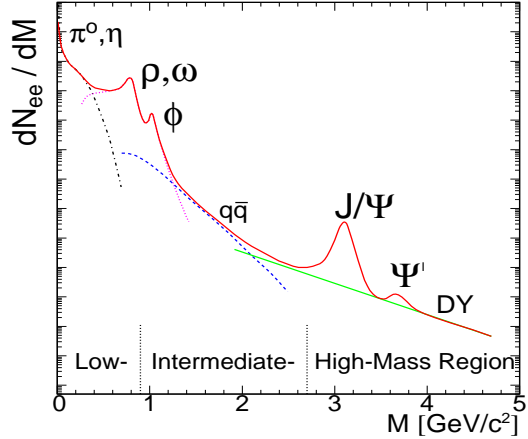,width=9.5cm,height=8.5cm}
\end{center}
\vspace{-1.5cm}
\caption{Schematic dilepton spectrum in ultrarelativistic heavy-ion
collisions.}
\label{fig_dilep}
\end{figure}

%%%%%%%%%%%%%%%%%%%%%%%%%%%%%%%%%%%%%%%%%%%%%%%%%%%%%%%%%%%%%%%%%%%%%%%%
Photons and dileptons emerging from heavy-ion collisions can be roughly
classified into three categories: (a) ``prompt" production associated 
with primordial $N$-$N$ collisions; (b) ``radiation" due to 
(multiple) reinteractions; (c) final state decays of produced 
hadrons (``cocktail").
A schematic view of a resulting dilepton spectrum is shown in 
Fig.~\ref{fig_dilep} (adopted from Ref.~\cite{Drees96} in slightly
modified form).
At high masses the prompt yield due to Drell-Yan (DY) annihilation will 
dominate due to its power-law dependence on $M$ (rather than exponential 
characteristic for thermal radiation). On top of the DY continuum the
hidden heavy-flavor vector mesons ($\Psi$, $\Upsilon$) are situated;
they decay long after strong interactions have ceased (category (c)),
with their small number being compensated by large 
e.m.~branching ratios, e.g. $\Gamma_{e^+e^-}/\Gamma_{tot}$=6\% (2.5\%) 
for $J/\psi$ ($\Upsilon$), about $10^3$ larger than
for $\rho$ or $\omega$ mesons.
The typical window for thermal radiation (category (b))
is below $M\simeq3$~GeV. The main background is (i) correlated charm 
decays,  $D\bar D\to e^+ e^- \nu_e \bar\nu_e X$, at intermediate mass, 
1~GeV~$\le$$M$$\le$~3~GeV, and (ii) Dalitz decays of light mesons
($\pi^0,\eta,\eta'$$\to$$\gamma e^+e^-$, $\omega\to\pi^0e^+e^-$) at
low mass, $M\le 1$~GeV. 
In the following we will focus on identifying characteristic regimes 
of thermal radiation. 

The response of a strongly interacting system to an
e.m.~excitation can be quite generally expressed in terms of the 
pertinent (time-ordered) current correlation function (which, to lowest 
order in $\alpha_{\rm em}$, is just the retarded photon 
selfenergy)~\cite{MT85,Weld90}, 
\beq
\Pi_{\rm em}(q) = -i \int d^4x \ e^{iqx} \ 
\langle\langle {\cal T} j^\mu(x) j^\nu(0) \rangle\rangle \  , 
\eeq 
where the brackets denote averaging over the states of
the system. E.g., in the case of deep-inelastic scattering the incoming 
photon is spacelike ($q^2<0$), the
average is over the nucleon, and the imaginary part of the correlator, 
Im~$\Pi_{\rm em}$, is directly related to the structure functions of the 
nucleon. In heavy-ion collisions, one is interested in the {\em emitted}
thermal radiation, which can be either lightlike (real photons, $q_0=q$)
or timelike (dileptons, $M^2\equiv q^2>0$). The respective emission rates 
are given by
\bea
q_0 & \frac{dN_\gamma}{d^4xd^3q} &= -\frac{\alpha_{\rm em}}{\pi^2} \
       f^B(q_0;T) \  {\rm Im}\Pi_{\rm em}^T(q_0=q;\mu_B,T) \ , 
\label{Rphot}
\\
 & \frac{dN_{e^+e^-}}{d^4xd^4q} &= -\frac{\alpha_{\rm em}^2}{M^2\pi^3} \
       f^B(q_0;T) \  {\rm Im}\Pi_{\rm em}(M,q;\mu_B,T) \ , 
\label{Rdilep}
\eea
($f^B$: thermal Bose distribution).
It is very important to note that both photon and dilepton 
emission are described by the {\em same} function,  
Im$\Pi_{\rm em}$\footnote{In fact, $\Pi_{\rm em}$ also governs charge 
fluctuations via the e.m. susceptibility, obtained from the static 
spacelike limit, $\chi_{\rm em}$=$\Pi_{\rm em}(q_0$=0,$q$$\to$0), cf. 
Ref.~\cite{PRWZ02} for a recent discussion of this point.}. Theoretical 
models for photon and dilepton spectra in URHICs ought to satisfy this 
consistency constraint. 

Eqs.~(\ref{Rphot}) and (\ref{Rdilep}) are to leading order in the 
e.m.~coupling $\alpha_{\rm em}$, but exact in the strong interactions
encoded in $\Pi_{\rm em}$. The photon rate is 
${\cal O}(\alpha_{\rm em})$, while it is 
${\cal O}(\alpha_{\rm em}^2)$ for dileptons. From the experimental 
point of view this appears to be an advantage for photon production, but  
from the theoretical point of view it is not. This is so because, 
to leading order in the {\em strong} interactions, the photon rate
is ${\cal O}(\alpha_{s})$ whereas the dilepton rate is
${\cal O}(1)$, and thus, at sufficiently high mass, under
better theoretical control. Indeed, in the vacuum, Im$\Pi_{\rm em}$    
is nonzero only for $M^2 > 4m_\pi^2$, and can be 
determined from $e^+e^-$ annihilation into hadrons. At low masses, 
$M\le 1$~GeV, $j_{\rm em}^\mu$ can be rather accurately saturated 
by the light vector mesons $\rho$, $\omega$, and $\phi$,
whereas at high masses it becomes amenable to a perturbative 
description in terms of quark fields, resulting in   
\beq
{\rm Im} \Pi_{\rm em}^{\rm vac}(M) = \left\{
\begin{array}{ll}
 \sum\limits_{V=\rho,\omega,\phi} \left(\frac{m_V^2}{g_V}\right)^2 \
{\rm Im} D_V(M) & , \ M \le M_{dual}
\vspace{0.3cm}
\\
-\frac{M^2}{12\pi} \ (1+\frac{\alpha_s(M)}{\pi} +\dots)  \ N_c
\sum\limits_{q=u,d,s} (e_q)^2  & , \ M \ge M_{dual} \ .
\end{array}  \right.
\label{Piem}
\eeq
From the inclusive cross section for $e^+e^-\to hadrons$ one finds
$M_{dual}\simeq 1.5$~GeV.           

For time-integrated yields from a heavy-ion collision it is useful
to estimate which temperature regime is predominately probed
at given $M$ (or $q_0$), cf. also~\cite{Shu80}.
Let us first focus on the dilepton case
and perform spatial integrations over Eq.~(\ref{Rdilep}) to obtain
\bea
\frac{dN_{ee}}{dMd\tau} &=& 
\frac{M}{q_0} \int d^3x \ d^3q \ \frac{dN_{ee}}{d^4xd^4q}
\nonumber\\
 &\simeq & {\rm const} \  V_{FB}(T) \ \frac{{\rm Im}\Pi_{\rm em}(M;T)}{M}
           \int \frac{d^3q}{q_0} \ {\e}^{-q_0/T} 
\nonumber\\
 &\simeq & {\rm const} \  V_{FB}(T) \ \frac{{\rm Im}\Pi_{\rm em}(M;T)}{M^2}
           \ \e^{-M/T} \ (MT)^{3/2}  
\eea
For simplicity, we have employed a uniform 3-volume as well as 
a nonrelativistic approximation, $M\gg T$. To infer the temperature
dependence of the volume and proper time, we assume isentropic
expansion, i.e., the total entropy $S$=$sV_{FB}$ to be conserved. 
Then $V_{FB}(T)\propto T^{-n}$ with $n$=3 for an ultrarelativisitc ideal gas, 
(e.g., $s=d_{QG} (4\pi^2/90) T^3$ for massless quarks 
and gluons with degeneracy $d_{QG}$), whereas $n$$\simeq$4-5 for a 
resonance hadron gas. The dependence of proper time on $T$, 
$\tau \propto T^{-k}$ (or $d\tau\propto T^{-k-1} dT$), 
is sensitive to the expansion dynamics: for the purely longitudinal 
(1-D Bjorken) case, appropriate for early phases in URHICs, one has 
$s_0\tau_0=s\tau$ and thus 
$\tau \propto T^{-n}$; in the later stages the expansion is closer to 3-D,
implying $\tau \propto T^{-n/3}$.
Collecting the various powers one finds  
\bea
\frac{dN_{ee}}{dMdT} &\propto& T^{-n-k-1} \ \e^{-M/T} \ (MT)^{3/2} \ 
     \frac{{\rm Im}\Pi_{\rm em}(M;T)}{M^2}  
\nonumber\\
&\propto& {\rm Im}\Pi_{\rm em}(M;T) \ \e^{-M/T} \ T^{-5.5} \ , 
\eea
approximately holding for both early and later phases. To find the
regime of maximal emission one simply takes the temperature
derivative.

At large $M$, where ${\rm Im}\Pi_{\rm em}(M;T)$ is only weakly dependent 
on $T$ (corrections are of order ${\cal O}(\alpha_sT^2/M^2)$), one finds 
$T_{max} \simeq M/5.5$, e.g.,
at $M=2$~GeV one has $T_{max}$=360~MeV, well inside the QGP. In fact, this 
estimate has to be taken with care, as the function $f(z)=\e^{-1/z}/z^{5.5}$ 
($z$$\equiv$$T/M$) has a substantial tail towards large $z$. Calculating 
the yield as a function of initial temperature $T_0$, 
$Y(z_0) \equiv \int\limits_{0}^{z_0} f(z) dz$,  
it turns out that 2/3 (90\%) of the limiting yield, $Y$($z_0$=$\infty$), 
is reached only for $z_0$$\simeq$0.3 (0.5). For $M$=2~GeV this implies 
$T_0$$\simeq600$~MeV (1~GeV), well above 
$T_{max}$. In practice this means: 
(i) at both RHIC and LHC dilepton radiation in the intermediate mass region 
is dominated by QGP emission, 
(ii) whereas at RHIC the yield is sensitive to the initial 
conditions, it could be close to its maximal value at LHC.    

An interesting opportunity could arise in the $M$$\simeq$1.5~GeV region 
through resonance states in the QGP as predicted by recent lattice QCD
calculations~\cite{AH03}. If the $\rho$ meson sur\-vives, a dilepton 
excess could occur~\cite{CS04} around its expected mass, 
$m_\rho\simeq 2m_q^{th}\sim 2gT$.
   
In the low-mass region, $M\le 1$~GeV, the situation is more complicated
since the e.m. correlator is directly proportional to the vector-meson 
spectral functions, which, via their in-medium modifications, induce 
further temperature- and density-dependencies (see also 
Sec.~\ref{ssec_lm} below).   
%Neglecting those, and fixing for definiteness $M$=0.5~GeV, one has 
%$T_{max}$$\simeq$90~MeV, with the same bias towards higher temperatures as 
%elucidated above. Taking as another extreme the in-medium spectral functions
%to be proportional to the total hadron density, $\rho_{had}\simeq N_{had}/V_{FB}$
%with constant $N_{had}$ (due to a chemical freezeout close to the beginning
%of the hadronic phase), one finds $T_{max}\simeq M$, which is well above
%$T_c$ so that the estimate is no longer applicable (many-body effects at 
%high-density will severely limit the increase in the spectral strength, 
%e.g., due to normalization conditions to be satisfied by the energy-integrated 
%spectral function). 
Model calculations~\cite{RW00} suggest that the largest contribution to 
the thermal yield below the free $\rho$/$\omega$ mass originates 
from the {\em hadronic} phase close to $T_c$, thus providing favorable 
conditions for probing chiral symmetry restoration.

Arguments similar to the dilepton case apply to photon radiation, where
the measured spectra are usually quoted as an invariant yield
$q_0 dN_\gamma/d^3q = dN_\gamma/dy d^2q_t$ versus transverse
momentum $q_t$ in a given range of rapidity, $y$. The temperature profile 
of the yield can be inferred as above, with 2 modifications:
(i) there is no integration over $q$, so no factor $(MT)^{3/2}$ arises;
(ii) even to lowest order in $\alpha_s$, Im$\Pi_{\rm em}$($q_0$=$q$)
   carries a leading temperature dependence, e.g. $\propto T^2$ in 
   perturbative QCD calculations~\cite{Shu78,KM81,KLS91,Baier92,AMY01}.    
Thus, at sufficiently high energies and for 1-D expansion one has
\beq
\frac{dN_\gamma}{dy d^2q_t} 
 \ \propto \  T^{-n-k-1} \ \e^{-q_0/T} \ {\rm Im}\Pi_{\rm em}(q_0=q) 
 \ \propto \  T^{-5} \ \e^{-q_0/T} \ , 
\eeq
which is almost identical to dileptons, i.e., photon radiation
at energies $q_0\ge 2$~GeV is also a sensitive probe of the early (QGP) 
phases.

At low energies, $q_0\le 1$~GeV, photon spectra are dominated by 
hadronic emission, which is again difficult to quantify due to the 
$T$-dependence of the e.m. correlator. 
Another subtlety for photons is that, 
unlike the dilepton invariant mass, the photon transverse momentum is 
Lorentz-variant and therefore sensitive to transverse flow velocities, 
entailing blue-shifts in the spectrum (more pronounced for the later stages). 

The main features of the above discussion are summarized
in Table~\ref{tab_pillar}. 
%One may add that, due to the somewhat greater 
%sensitivity of the e.m. correlator to temperature for $q_0=q$, the thermal 
%photon yield will be somewhat more biased towards higher temperatures
%than dileptons (which is, of course, related to the fact that the photon
%rate is ${\cal O}(\alpha_s)$).
\begin{table}[!t]
\begin{center}
%\hskip4pc\vbox{\columnwidth=26pc
\begin{tabular}{c|c|c}
          & Low Mass/Energy  & Intermediate Mass/Energy \\
\hline
              &  in-medium $\rho$, $\omega$, $\phi$  
                    & continuum emission ($q\bar q\to ee$)       \\
  Dileptons  &  Chiral Symmetry  &  leading order ${\cal O}(\alpha_{\rm em}^2)$\\ 
              &    Restoration?     &    QGP Radiation?\\
\hline
               &  hadron decays/scattering  &  continuum emission \\
Photons    &  $a_1\to\pi\gamma$, $\pi\rho\to\pi\gamma$    
                     & leading order ${\cal O}(\alpha_{\rm em}\alpha_s)$  \\
           & Medium Effects?  & QGP Radiation? \\ 
\end{tabular}
%}
\end{center}
\caption{The four pillars of thermal emission of electromagnetic radiation.}
\label{tab_pillar}
\end{table}

%%%%%%%%%%%%%%%%%%%%%%%%%%%%%%%%%%%%%%%%%%%%%%%%%%%%%%%%%%%%%%%%%%%%%%%%
%\subsection{Quarkonia}
%\label{ssec_quarkonia}
%%%%%%%%%%%%%%%%%%%%%%%%%%%%%%%%%%%%%%%%%%%%%%%%%%%%%%%%%%%%%%%%%%%%%%%%
%Although not included in the oral presentation at the meeting, let us make
%a few remarks on heavy quarkonium physics in URHICs.
%In recent years, the original suggestion of charmonium suppression 
%due to Debye screening as a signature of QGP formation~\cite{MS86} has 
%evolved into a more complex subject, especially at collider energies. 
%The main new ingredient is regeneration of charmonium states at the
%phase boundary~\cite{GG99,pbm00,Goren01,GR01} or through the backward 
%channel of dissociation reactions~\cite{Thews01,Zhang02,GRB04}.     
%
%relation to deconfinement not obvious
%test resonances observed in lattice QCD
%strong increase of width across $T_c$?

%%%%%%%%%%%%%%%%%%%%%%%%%%%%%%%%%%%%%%%%%%%%%%%%%%%%%%%%%%%%%%%%%%%%%%%%
\section{Thermal Dileptons}
\label{sec_dilep}
%%%%%%%%%%%%%%%%%%%%%%%%%%%%%%%%%%%%%%%%%%%%%%%%%%%%%%%%%%%%%%%%%%%%%%%%

%%%%%%%%%%%%%%%%%%%%%%%%%%%%%%%%%%%%%%%%%%%%%%%%%%%%%%%%%%%%%%%%%%%%%%%%
\subsection{Chiral Symmetry}
\label{ssec_chiral}
%%%%%%%%%%%%%%%%%%%%%%%%%%%%%%%%%%%%%%%%%%%%%%%%%%%%%%%%%%%%%%%%%%%%%%%%
Let us recollect some elements of chiral symmetry, its breaking
and restoration, which are relevant for the subsequent discussion.

\begin{figure}[!t]
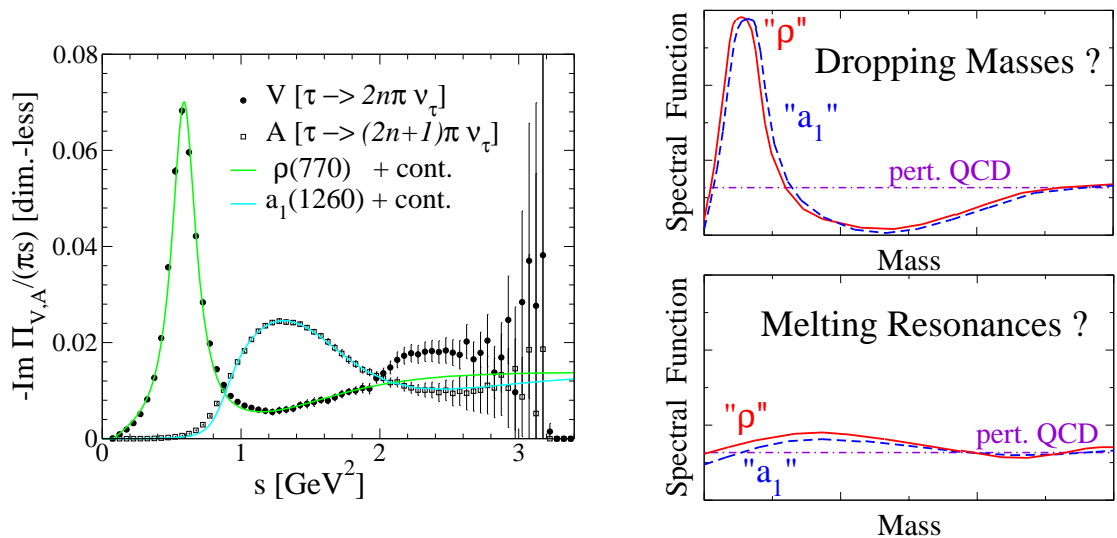

%\vspace{-0.8cm}
%\hspace{-0.5cm}
\begin{minipage}{7.5cm}
\epsfig{file=VAfit3.eps,width=7.5cm}
\end{minipage}
\hspace{1cm}
\begin{minipage}{7cm}
\epsfig{file=BR2-scenario.eps,width=6cm}
\epsfig{file=RW2-scenario.eps,width=6cm}
\end{minipage}
%\vspace{-0.5cm}
\caption{Left panel: vector and axialvector spectral functions as measured in
hadronic $\tau$ decays~\cite{aleph98} with model fits using vacuum $\rho$ and
$a_1$ spectral functions plus perturbative continua~\cite{Rapp02};
right panel: two possible scenarios for chiral symmetry restoration
in matter.}
\label{fig_VA}
\end{figure}

For the two lightest quark flavors ($u$ and $d$) the QCD Lagrangian 
\beq
{\cal L} = \bar q (i\not{\!\!D} - \hat m_q) q -\frac{1}{4} (G_a^{\mu\nu})^2  
\eeq
possesses a $SU(2)_L \times SU(2)_R$ (``chiral") symmetry (a combination of 
invariance in isospin and left- and right-handed quark-fields). It 
is only slightly broken explicitly by small {\em current} quark masses
$m_{u,d}\simeq5-10$~MeV (which enter the matrix 
$\hat m_q=diag(m_u, m_d)$ and originate from electroweak 
interactions). A much more dramatic
phenomenon is the {\em spontaneous} breaking of chiral symmetry (SBCS), 
which is induced by a strong attraction in the scalar $q\bar q$ channel
leading to the formation of a quark condensate, $\langle 0 |\bar q q|0\rangle
\simeq -(250MeV)^3$, filling the QCD vacuum. Although the condensate
is not an observable, it manifests itself in (hadronic)
excitations of the vacuum, e.g. 
(i) at the quark level, an energy gap 
    $\Delta_{\bar qq}\equiv m_q^*\propto \langle \bar qq\rangle$
    generates a {\em constituent} quark mass $m_q^*\simeq 350$~MeV which 
    constitutes the major portion of the visible mass in the universe;
(ii) there appear 3 (almost) massless (quasi-) Goldstone bosons, the pions 
     ($m_\pi^2\propto m_q$); 
(iii) a substantial mass difference of $\Delta M\simeq 0.5$~GeV 
      splits hadronic states within chiral multiplets (which in a chirally 
      symmetric vacuum state would be degenerate).     
For the light vector mesons in the 2-flavor sector, the $\omega(782)$
turns out to be a chiral singlet, whereas the chiral partner of
the $\rho$ meson is usually identified with the $a_1(1260)$\footnote{In
recent work by Harada \etal~\cite{HY01} the ``vector manifestation" of 
chiral symmetry has been suggested in which the chiral partner of the
(longitudinal component of the) $\rho$ meson is identified with the 
pion.}. 
The respective vector ($V$) and axialvector ($A$) spectral functions 
have been measured in $\tau$ decays, cf. left panel of Fig.~\ref{fig_VA}. 
Albeit limited by the $\tau$ mass, 
$m_\tau^2$=3.16~GeV$^2$, the spectra indicate the approach
to the perturbative limit (and each other) at large $s$=$q^2$, 
-Im$\Pi_{{\rm em},pert}^{I=1}/(\pi s)$=$\frac{N_c}{12\pi^2}$$\times$$\frac{1}{2}$, 
cf.~Eq.~(\ref{Piem}) (the factor of 
$\frac{1}{2}$ counts the isospin $I$=1 part of the correlator, with 
the remaining $\frac{1}{18}$ supplemented by the $I$=0  
$\omega$(782)), affirming that SBCS is a 
low-energy phenomenon. The connection to SBCS can be     
quantified by the 2.~Weinberg sum rule~\cite{Wein67},
\beq
f_\pi^2 = - \int \frac{ds}{\pi s} \
      \left[ {\rm Im} \Pi_V(s) -{\rm Im} \Pi_A(s) \right]  \ ,  
\label{wsr2}
\eeq 
relating the pion decay constant $f_\pi$, one of the order parameters 
of SBCS (the ``pole strength" of the Goldstone mode), to the 
(integrated) difference of 
%the $1/q^2$-weighted 
$V$ and $A$ spectral functions. Note that it is not only the mass, but 
the entire spectral shape that matters. Importantly, Eq.~(\ref{wsr2})
remains valid at finite temperature~\cite{KS94}, replacing 
$s$$\to$$q_0^2$ at fixed $q$.

Towards the critical temperature ($T_c$), chiral symmetry restoration requires
the $V$ and $A$ spectral functions to (approximately) degenerate (lattice 
calculations show that $\langle \bar qq\rangle(T)/\langle \bar qq\rangle(0)$ 
drops rather rapidly around $T_c$ down to $\sim$10-15\% slightly above
$T_c$). {\em How} this is realized, is one of the main questions 
in strong interactions, shedding light on the question
of mass generation. Two of the infinitely many possibilities are
illustrated in Fig.~\ref{fig_VA} (right panel).

%%%%%%%%%%%%%%%%%%%%%%%%%%%%%%%%%%%%%%%%%%%%%%%%%%%%%%%%%%%%%%%%%%%%%%%%
\subsection{Low-Mass Dileptons}
\label{ssec_lm}
%%%%%%%%%%%%%%%%%%%%%%%%%%%%%%%%%%%%%%%%%%%%%%%%%%%%%%%%%%%%%%%%%%%%%%%%
Nature is kind to us in that it rather directly lets us probe the isospin-1
part of the vector correlator:  in the thermal dilepton rate, 
Eq.~(\ref{Piem}), the isovector (``$\rho$") channel dominates 
over the isoscalar (``$\omega$") by a factor 
$\Gamma_{\rho\to ee}/\Gamma_{\omega\to ee}$=$g_\omega^2/g_\rho^2$$\simeq$10.
Consequently, vigorous theoretical efforts  have been devoted
to assess medium modifications of the $\rho$ meson (see 
Ref.~\cite{RW00} for a review), especially after it became clear that 
CERES data~\cite{ceres160} are incompatible with a vacuum $\rho$-meson 
line shape. 

\begin{figure}[!t]
%\vspace{-0.8cm}
%\hspace{-0.5cm}
\begin{minipage}{7.5cm}
\epsfig{file=Arho672Ca.eps,width=6.7cm,height=5.3cm}
\end{minipage}
\hspace{-0.6cm}
\begin{minipage}{7cm}
\epsfig{file=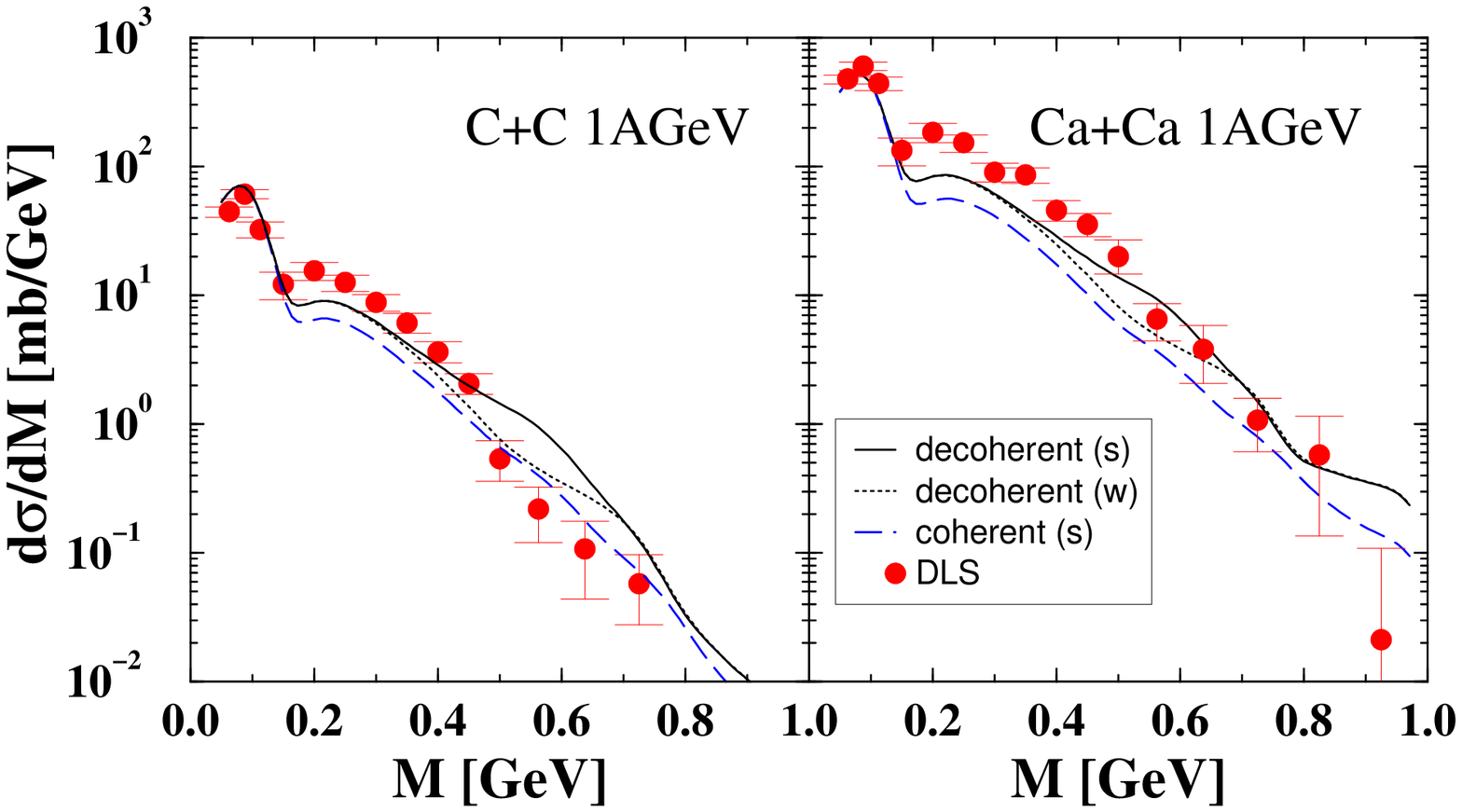,height=6cm,width=8.5cm}
\vspace{-1cm}
\end{minipage}
%\vspace{-0.5cm}
\caption{Left panel: $\rho$ spectral function under conditions
resembling heavy-ion collisions at BEVALAC/SIS energies~\cite{BCRW98}.
Right panel: transport calculations~\cite{Fuchs03}
of $e^+e^-$ cross sections in 1~AGeV $C$+$C$ and $Ca$+$Ca$ collisions 
employing broadened $\rho$ and $\omega$ spectral functions with 
(long-dashed line) and without (solid and dotted lines) coherence 
effects in baryon resonance decays, compared to DLS data~\cite{dls97}.}
\label{fig_dls}
\end{figure}

In hadronic many-body approaches one evaluates an in-medium 
$\rho$ propagator, 
\beq
D_\rho(M,q;\mu_B,T) = \left[ M^2-(m^{(0)}_\rho)^2-\Sigma_{\rho\pi\pi}
             -\Sigma_{\rho B} - \Sigma_{\rho M} \right]^{-1} \ ,  
\eeq 
in terms of selfenergies arising from direct interactions with surrounding
baryons ($\Sigma_{\rho B}$) and mesons ($\Sigma_{\rho M}$), as well as  
from (in-medium) $\pi\pi$ loops ($\Sigma_{\rho\pi\pi}$). Underlying
hadronic Lagrangians are constrained by free decay widths and/or 
scattering data (including pho\-toabsorption on nucleons and 
nuclei~\cite{RUBW98}), and resulting spectral functions in nuclear 
matter have been found to satisfy QCD sum 
rules~\cite{KKW97,Leup98}.     
The rather generic results of such calculations are (i) a substantial broadening
of the $\rho$ spectral function, with little mass shift (real parts of 
various contributions to $\Sigma_\rho$ tend to cancel, whereas imaginary
parts strictly add up), (ii) the prevalence of baryonic over mesonic effects 
(also confirmed in chiral expansion schemes~\cite{SZ99}), 
cf.~Fig.~\ref{fig_dls} (left panel) for an example.   

The ``dropping $\rho$-mass" scenario, which was originally suggested
within a mean-field approach exploiting scale invariance of (classical) 
QCD~\cite{BR91}, has received renewed support in the 
``vector manifestation" of chiral symmetry~\cite{HY01} (see footnote above):
at $T_c$ the longitudinal component of the $\rho$ (rather than the 
scalar ``$\sigma$" field) degenerates with the pion, forcing its
bare mass $m_\rho^{(0)}$ to (approximately) zero.  
 
Concerning applications to heavy-ion collisions, let us start from
low energies. Current calculations cannot account for the large 
enhancement observed by DLS in $C$+$C$ and $Ca$+$Ca$ collisions at 
1~AGeV, neither with many-body spectral functions~\cite{BCRW98} nor
with a dropping $\rho$-mass, nor with both~\cite{BK99}. More recently
it has been pointed out~\cite{Fuchs03} that a decoherence of the 
lepton-pair producing sources (i.e. virtual vector mesons implemented 
with destructive phase factors in elementary $p$-$p$ collisions)
in dense matter can lead to some additional (albeit not enough) 
enhancement in the DLS spectra, cf. right panel of Fig.~\ref{fig_dls}. 
At the same time, optimal (fitted) values for in-medium $\rho$ and 
$\omega$ widths have been extracted~\cite{Fuchs03} which are consistent 
with the many-body calculations discussed above (left panel of 
Fig.~\ref{fig_dls}).  New precision data in this energy regime
are expected soon from the HADES experiment~\cite{HADES} at SIS (GSI).

\begin{figure}[!t]
%\vspace{-0.8cm}
%\hspace{-0.5cm}
\begin{minipage}{7.7cm}
\epsfig{file=dlsQH69PbyT250-956.eps,width=7.7cm,height=6.4cm}
\end{minipage}
\hspace{-0cm}
\begin{minipage}{8cm}
\epsfig{file=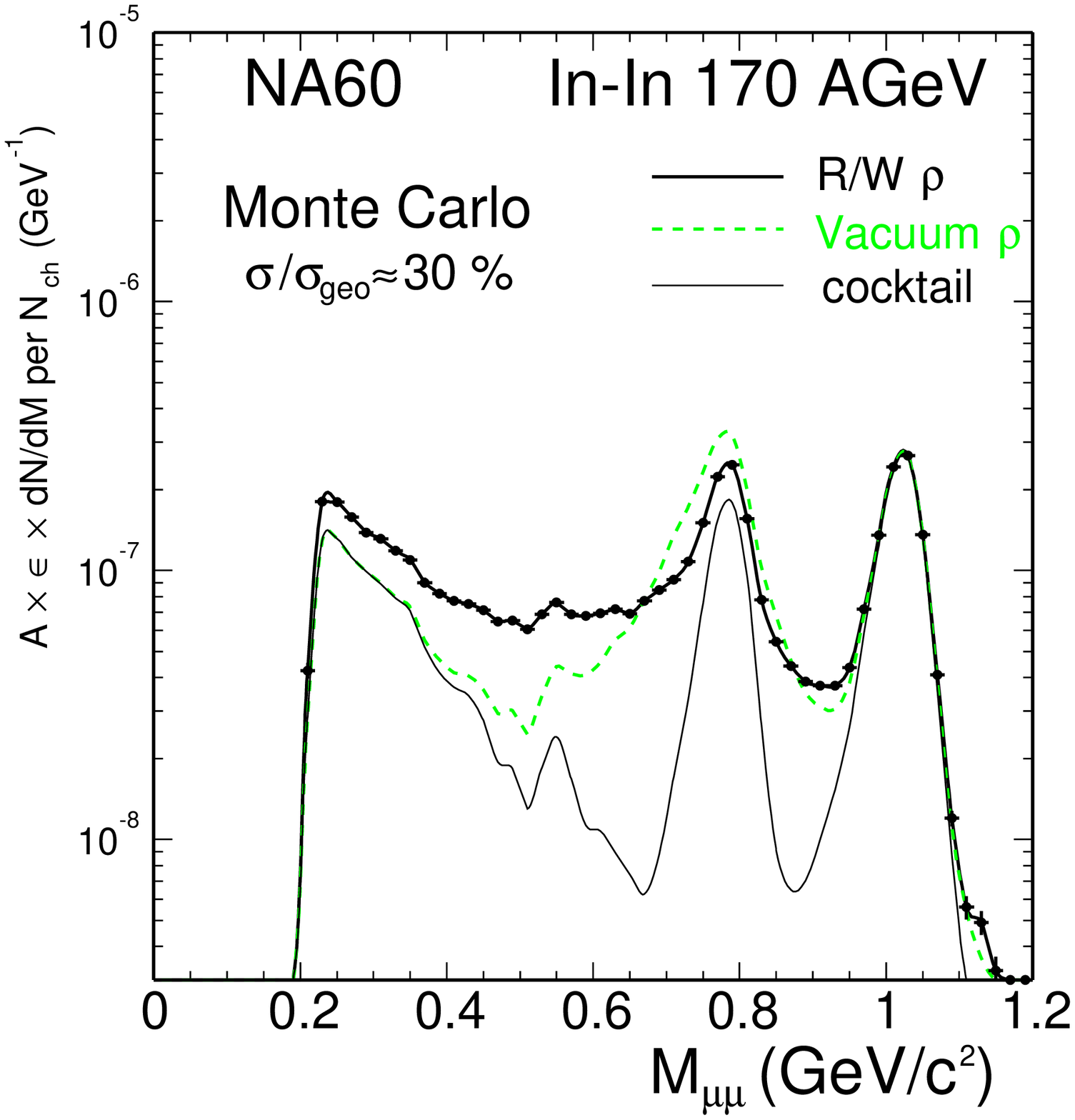,width=8cm,height=6.3cm}
\vspace{-1cm}
\end{minipage}
%\vspace{-0.5cm}
\caption{Left panel: $e^+e^-$ spectra in 
$Pb$(158AGeV)+$Au$~\cite{ceres160} compared to thermal fireball  
calculations~\cite{RW99,Rapp02} using free (dotted line), 
in-medium many-body (solid line) and dropping-mass $\rho$-spectral functions. 
Right panel: {\em Simulations} of $\mu^+\mu^-$ spectra~\cite{Sanja03}
as expected to be measured by NA60 in $In$(170AGeV)+$In$~\cite{na60} 
(solid curve: hadron decays, dashed curve and full circles: using free 
and in-medium~\cite{RW99} Im$D_\rho$, respectively.}
\label{fig_sps}
\end{figure}
At SPS energies, theoretical models compare more favorable with existing
data, see left panel of Fig.~\ref{fig_sps}.
However, at this point both hadronic many-body calculations and a 
dropping $\rho$-mass scenario are compatible with the CERES 
data~\cite{ceres160}. The prevalence of baryon-driven medium effects 
predicted within the many-body approach~\cite{RCW97,RW99}
has recently been confirmed experimentally by a relatively larger 
excess observed at lower SPS energy of 40~AGeV~\cite{ceres40}.
Also note that the QGP contribution to the in-medium yield in the 
low-mass region is at the 10-15\% level.
Very promising new data are expected from the NA60 experiment at 
SPS~\cite{na60}. The larger available data sample from muon pairs 
combined with an improved low-$q_t$ capability due to a new vertex 
detector leads to simulation results~\cite{Sanja03} indicating excellent 
resolution and statistics, cf. right panel of 
Fig.~\ref{fig_sps}. In particular $\omega$- and $\phi$-peaks will be 
clearly discernible.

In-medium effects on the $\omega$ meson~\cite{KKW97,Ra01,PM01,RK04,ME04}
are expected to be of si\-milar magnitude as for the $\rho$, but the 
$\phi$ seems to be more robust (which is probably related to the 
OZI rule suppressing resonant $\phi$-N interactions). 
Nevertheless, there are interesting issues related to the $\phi$-meson, 
e.g. comparisons of its line shape and yield in $l^+l^-$ vs. $K^+K^-$
decay channels, see, e.g., Haglin's talk at this meeting~\cite{Hag04}. 

%%%%%%%%%%%%%%%%%%%%%%%%%%%%%%%%%%%%%%%%%%%%%%%%%%%%%%%%%%%%%%%%%%%%%%%%
\subsection{Intermediate Mass Dileptons}
\label{ssec_im}
\begin{figure}[!tb]
\begin{minipage}{7cm}
\epsfig{file=dlPb38031na50.eps,width=7cm}
\end{minipage}
\hspace{0.5cm}
\begin{minipage}{7.5cm}
\epsfig{file=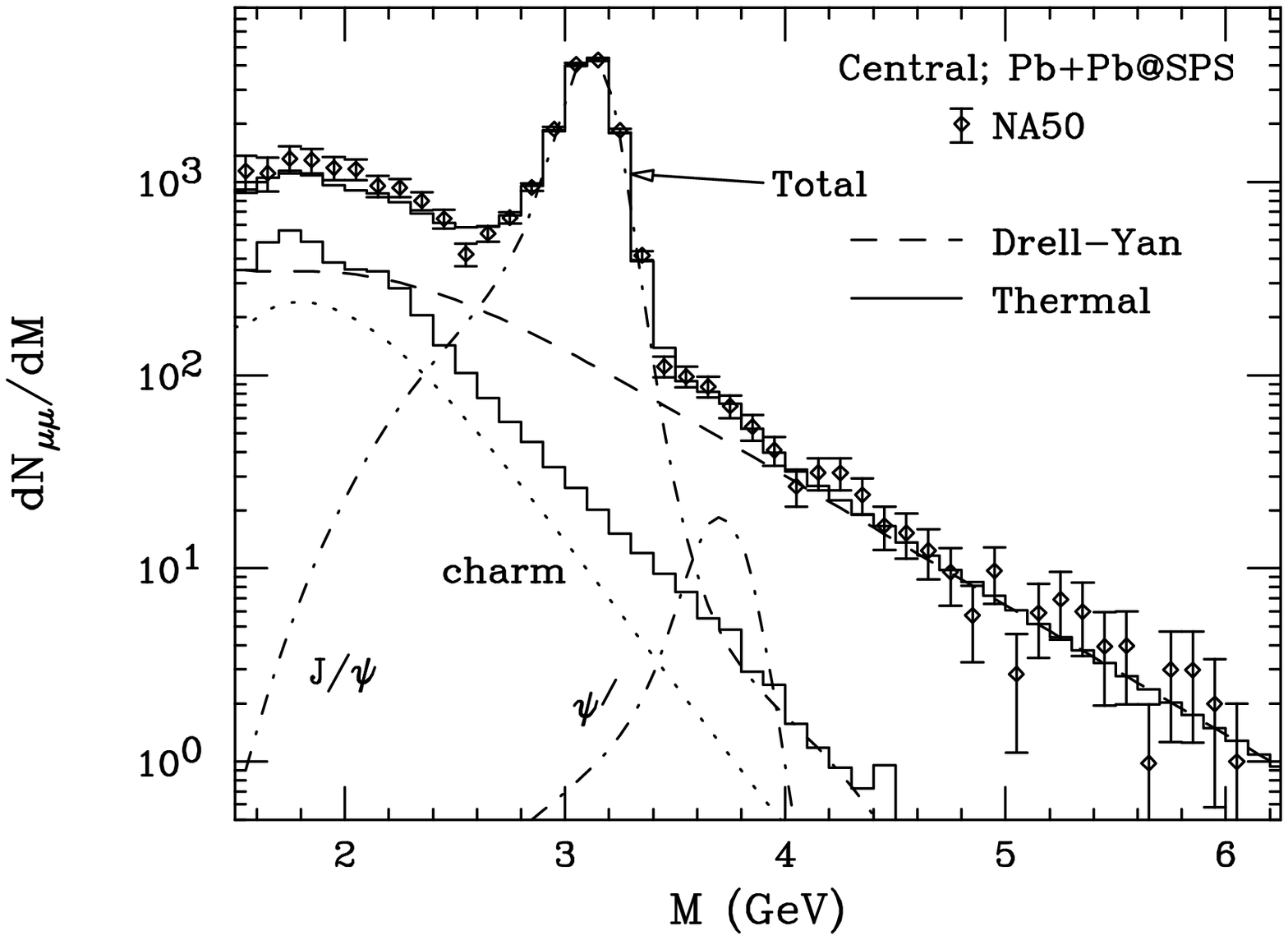,width=7.5cm}
\end{minipage}
%\vspace{-0.5cm}
\caption{$\mu^+\mu^-$ spectra measured by NA50~\cite{na50} at the SPS
compared to theoretical calculations addressing the excess at 
intermediate masses, 1.5~GeV~$\le M_{\mu\mu}\le$~3~GeV. Left panel:
thermal fireball model~\cite{RS00}; right panel: hydrodynamical 
model~\cite{KGS02}.}
\label{fig_im}
\end{figure}

%%%%%%%%%%%%%%%%%%%%%%%%%%%%%%%%%%%%%%%%%%%%%%%%%%%%%%%%%%%%%%%%%%%%%%%%
The main issue at dilepton masses above $\sim$1.5GeV is how
the QGP signal compares to competing sources, i.e., Drell-Yan 
annihilation, hadron gas radiation and correlated open-charm decays. 
Assuming the emission rate to be given by its perturbative form, 
Eq.~(\ref{Piem}), the key ingredient is the space-time evolution of the 
system. In Fig.~\ref{fig_im} a hydrodynamic~\cite{KGS02} (right panel) 
and a thermal fireball calculation~\cite{RS00}\footnote{The same 
evolution is underlying the results shown in Figs.~\ref{fig_sps} and
\ref{fig_phot}.} 
(left panel) are compared to NA50 data~\cite{na50} from central 
$Pb$(158~AGeV)-$Pb$ collisions at SPS.
Whereas in Ref.~\cite{RS00} 2/3 of the thermal yield 
originates from the hadronic phase (1/3 from a QGP with uniform 
initial temperature of
$T_0$$\simeq$210~MeV), Ref.~\cite{KGS02} assigns a
larger fraction to the QGP (induced by initial temperatures
$T_0$$\simeq$300~MeV). This difference deserves further investigation;
two possible reasons are:
(i) the hydrodynamic equations are solved assuming boost invariance
which is not present in the fireball parametrization, (ii) the hadronic
equation of state is in chemical equilibrium in the hydro-calculation
while in the fireball expansion meson-chemical potentials are
introduced to conserve the hadron ratios after chemical freezeout.
Nevertheless, in either case the thermal yield accounts  
for the observed excess, which, in particular, leaves little room
for an ``anomalous enhancement" of open-charm production. Also 
in this context, further experimental scrutiny is expected
from upcoming NA60 data~\cite{na60}.

%%%%%%%%%%%%%%%%%%%%%%%%%%%%%%%%%%%%%%%%%%%%%%%%%%%%%%%%%%%%%%%%%%%%%%%%
\subsection{Prospects for RHIC and Future Developments}
\label{ssec_dlrhic}
\begin{figure}[!tb]
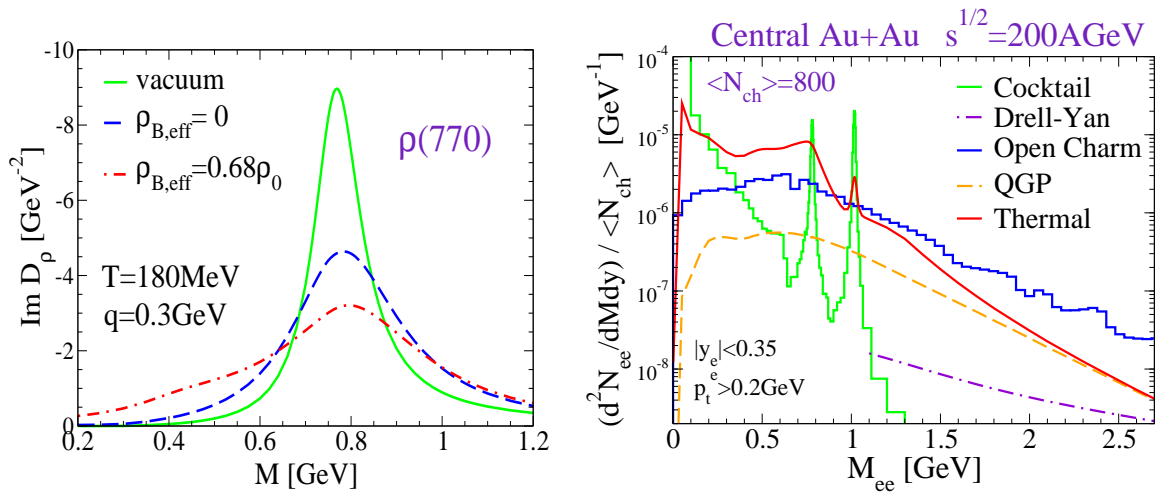

\vspace{0.3cm}
\begin{minipage}{7cm}
\epsfig{file=ArhoTr180.eps,width=7.2cm,height=6cm}
\end{minipage}
\hspace{0.4cm}
\begin{minipage}{7cm}
\vspace{-0.4cm}
\epsfig{file=dNdMAu200C.eps,width=7.6cm,height=6.45cm}
\end{minipage}
\caption{Left panel: $\rho$-meson spectral function in vacuum and under
RHIC con\-ditions with (dash-dotted line) and without (dashed line) the 
effects of anti-/baryons~\cite{Ra01}. Right panel: $e^+e^-$ spectra 
at RHIC; histograms: final-state decays of open-charm and light 
hadrons~\cite{Aver01}, solid line: combined thermal yield~\cite{Ra01} 
from hadronic matter (using in-medium vector spectral functions) and QGP 
(dashed line).}
\label{fig_rhic}
\end{figure}

%%%%%%%%%%%%%%%%%%%%%%%%%%%%%%%%%%%%%%%%%%%%%%%%%%%%%%%%%%%%%%%%%%%%%%%%
At RHIC, both low- and intermediate-mass dileptons will
be measured by PHENIX~\cite{Nagle03}. 
For medium effects on the low-mass vector mesons it is important
to realize~\cite{Ra01} that the relevant quantity is not the 
{\em net} baryon density (which is small at RHIC), but the {\em total}, 
$\varrho_{tot}\equiv\varrho_B+\varrho_{\bar B}$ (due to $CP$-invariance
of strong interactions mesons interact equally with 
baryons and antibaryons).
The combined effect of $B$ and $\bar B$ on the $\rho$ spectral function
at RHIC is indeed substantial, especially at masses below 0.5~GeV, 
see Fig.~\ref{fig_rhic} (left panel).   
Quantitatively also important is the conservation of the $\bar B$-number
in the hadronic evolution subsequent to chemical freeze\-out~\cite{Ra02},
which is necessary to maintain the observed hadron ratios (and implies a
significant $B$+$\bar B$ density, and thus stronger medium effects,
in the later phases). The ensuing (space-time integrated) thermal
dilepton spectrum~\cite{Ra01} in central $Au$-$Au$ collisions (right
panel of Fig.~\ref{fig_rhic}) exhibits an essentially melted $\rho$
resonance, while the $\omega$ and $\phi$ resonance regions are mostly
populated by the hadronic cocktail~\cite{Aver01} (i.e., decays after 
freezeout).

At intermediate masses, $M_{ee}>1.5$~GeV, the thermal yield is
dominated by QGP radiation. However, the increase
in $c\bar c$ production by about a factor $\sim$100 over the yield at
SPS renders correlated charm decays~\cite{Aver01} the prevalent 
source, at least if their transverse-momentum distributions are
taken from production in hard (primordial) $N$-$N$ collisions.
If, on the other hand, $c$-quarks undergo re-interactions
(e.g.~within the QGP), the isotropization of their momentum 
distributions is likely to lead to a softening of the pertinent dilepton
invariant-mass spectra, possibly re-opening the window on QGP radiation. 
Due to transverse-flow effects on the relatively heavy $c$-quarks this 
softening will be less pronounced than originally expected~\cite{Shu97}.   
        
Future developments from the theoretical side will have to
address the question of in-medium $a_1$ spectral functions, to
strengthen the connections to chiral symmetry restoration (it may
even be possible to study medium modifications the $a_1$ experimentally
via $\pi\gamma$ invariant-mass spectra). 
The importance of baryon effects in the $\rho$ and $\omega$ propagators
calls for a more detailed analysis of baryon properties
themselves in hot hadronic matter, see van Hees' talk at this 
meeting~\cite{Hees04b}.

Furthermore, hadronic models be can used to calculate Euclidean 
correlation functions via a straightforward folding with a 
thermal factor,   
\beq
\Pi_\alpha(\tau,q;T)=\int\limits_0^\infty dq_0 \ {\rm Im}\Pi_\alpha(q_0,q)
\ \frac{\cosh[q_0(\tau-1/2T)]}{\sinh[q_0/2T]}
\eeq
($\alpha=V, A$), which provides a direct means of comparison to 
lattice ``data" for $\Pi_\alpha(\tau)$, rather than having to apply an 
inverse integral transform to the latter.

%%%%%%%%%%%%%%%%%%%%%%%%%%%%%%%%%%%%%%%%%%%%%%%%%%%%%%%%%%%%%%%%%%%%%%%%
\section{Thermal Photons}
\label{sec_phot}
%%%%%%%%%%%%%%%%%%%%%%%%%%%%%%%%%%%%%%%%%%%%%%%%%%%%%%%%%%%%%%%%%%%%%%%%

%%%%%%%%%%%%%%%%%%%%%%%%%%%%%%%%%%%%%%%%%%%%%%%%%%%%%%%%%%%%%%%%%%%%%%%%
%\subsection{Production Rates in QGP and HG}
%\label{ssec_rates}
%%%%%%%%%%%%%%%%%%%%%%%%%%%%%%%%%%%%%%%%%%%%%%%%%%%%%%%%%%%%%%%%%%%%%%%%
Recent progress in assessing thermal photon rates from hot and dense
matter has been reviewed by several 
authors~\cite{Alam01,PT02,GH03,Ra04}.  As emphasized in 
Sec.~\ref{sec_emrad}, one of the diffi\-culties is that the leading 
term in the thermal photon rate is to nontrivial order in $\alpha_s$, 
implying the absence of a vacuum baseline
for Im$\Pi_{\rm em}$ at the photon point. 

\begin{figure}[!t]
\vspace{0.2cm}
\begin{minipage}{7.2cm}
\epsfig{file=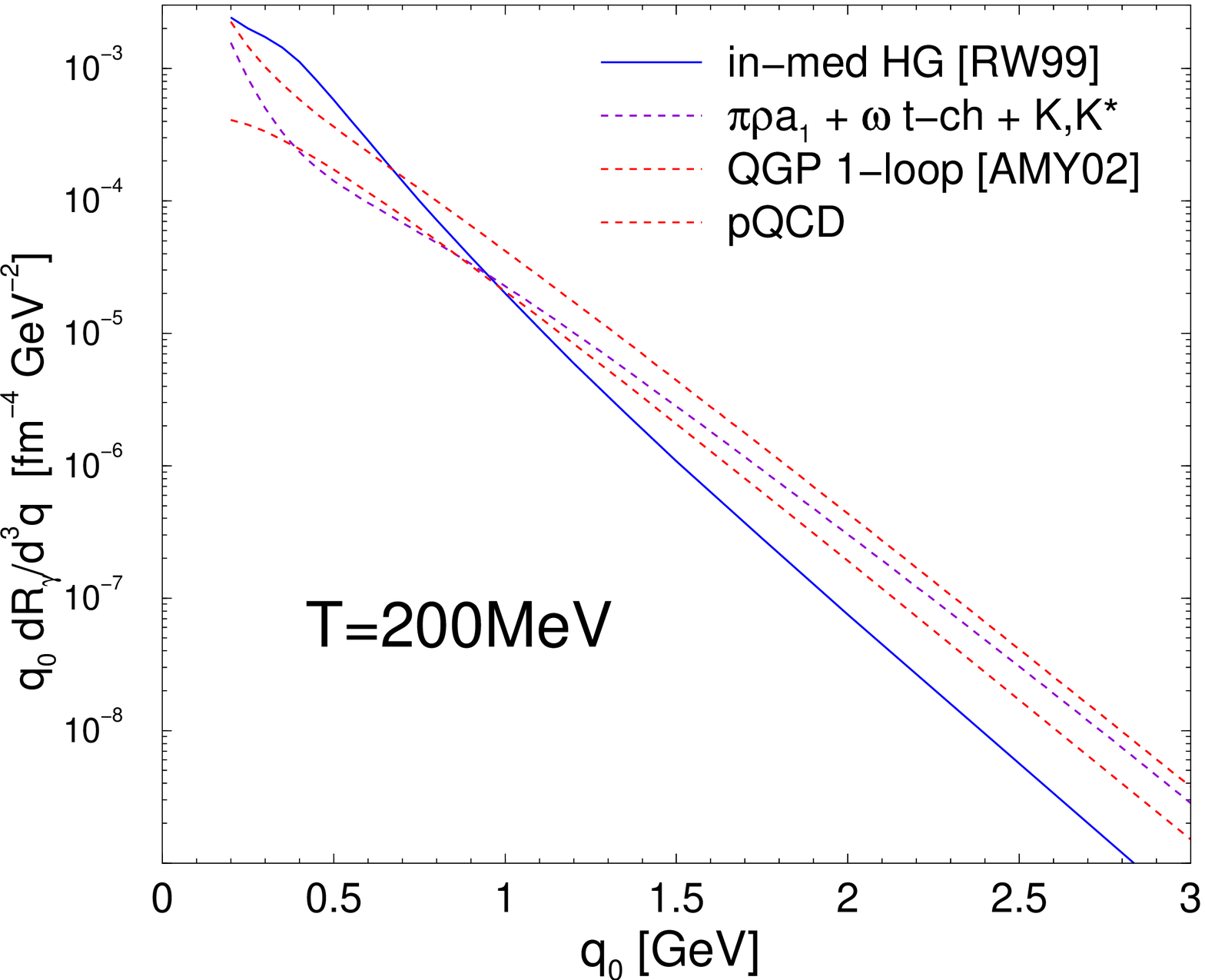,width=7.2cm,height=6cm}
\end{minipage}
\hspace{0.7cm}
\begin{minipage}{7.2cm}
\vspace{-0.5cm}
\epsfig{file=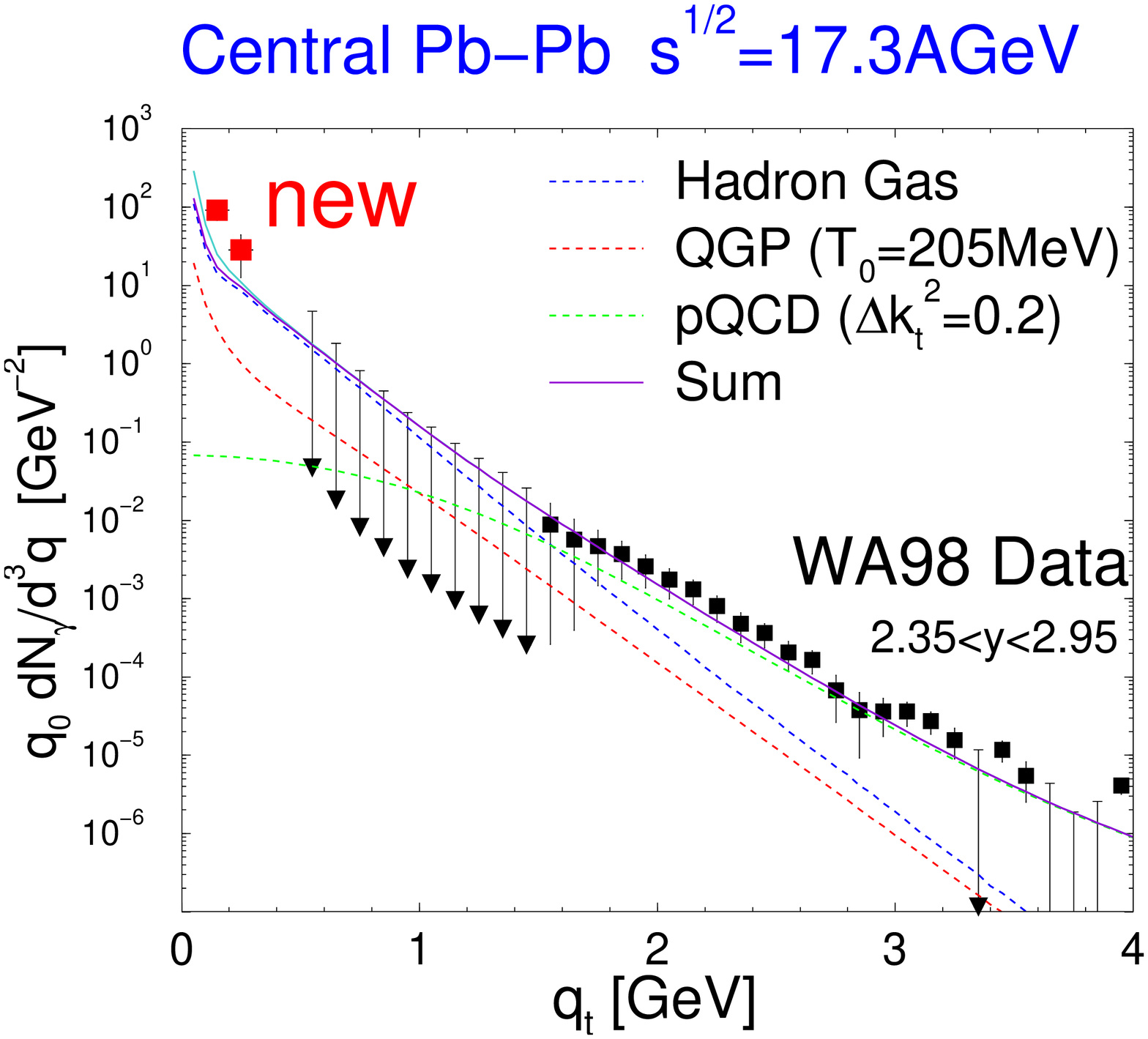,width=7.2cm,height=6.7cm}
\end{minipage}
\caption{Left panel: Thermal photon rates from hot hadronic matter~\cite{TRG04}
(solid and long-dashed line) and QGP (short-dashed line: tree-level
pQCD~\cite{KLS91,Baier92}, dashed-dotted line: complete leading
order~\cite{AMY01}. Right panel: WA98 data~\cite{wa98-00,wa98-04} 
compared to thermal fireball calculations~\cite{TRG04}; the upper solid
line at low $q_t$ additionally includes Bremsstrahlung from  
$S$-wave $\pi\pi$ scattering~\cite{TRG05}.
}
\label{fig_phot}
\end{figure}

For the QGP, it has been realized~\cite{AGKZ98} that early perturbative 
(tree-level) calculations~\cite{Shu78,KM81,KLS91,Baier92} for 
$q\bar q\to g\gamma$ and $qg\to q\gamma$ receive contributions at the
same (leading) order from Bremsstrahlung's processes involving $t$-channel
gluon exchange, due to forward infrared singularities in the latter.
The required resummation to obtain the full result has been accomplished
in Ref.~\cite{AMY01}; the pertinent rates exhibits a factor 2-3 
enhancement over the early results, cf. Fig.~\ref{fig_phot} 
(left panel). For hot hadronic matter, $t$-channel meson exchange is 
expected to be the predominant source of high-energy photons, most 
notably $\pi$ exchange in $\pi\rho\to\pi\gamma$. There has been some 
controversy in the 
literature concerning the role of the $a_1$ $s$-channel contribution
(or, equivalently, $a_1$ decays): on general grounds, this source should 
be suppressed at high energy due to an extra $1/s$-dependence in the $a_1$
propagator, which is even more pronounced if hadronic vertex form factors
are included based on a $\chi^2$-fit to the $a_1$ hadronic 
and radiative decay branchings in vacuum~\cite{TRG04}.
Baryonic sources have recently been assessed in Refs.~\cite{Alam03,TRG04}.
In Ref.~\cite{TRG04}, previous calculations of the e.m. correlator in the 
timelike regime~\cite{RW99} were carried to the photon point, thus 
establishing consistency with in-medium dilepton rates. One finds that 
the baryonic effects in the
photon rate dominate hadronic emission for energies $\sim$0.2-1~GeV (cf. 
left panel of Fig.~\ref{fig_phot}). Another, somewhat surprising, result
of Ref.~\cite{TRG04} is that $\omega$ $t$-channel exchange in 
$\pi\rho\to\pi\gamma$ becomes the dominant reaction at high energies
(due to the large $\pi\rho\omega$ coupling constant).  Reactions 
involving strangeness were found to contribute at the 
$\sim$25\% level. 

A comparison between (top-down extrapolated) LO-QGP and (bottom-up 
extrapo\-lated) total hadronic rates in the vicinity of $T_c$ indicates 
that both are very similar  (this has also been noticed for dilepton
rates~\cite{RW99,Rapp02}). If not a coincidence, it could 
be related to a kind of quark-hadron duality, reminiscent to  
inclusive electron scattering~\cite{Nicu00}.  

%%%%%%%%%%%%%%%%%%%%%%%%%%%%%%%%%%%%%%%%%%%%%%%%%%%%%%%%%%%%%%%%%%%%%%%%
%\subsection{Comparison to Experiment}
%\label{ssec_sps}
%%%%%%%%%%%%%%%%%%%%%%%%%%%%%%%%%%%%%%%%%%%%%%%%%%%%%%%%%%%%%%%%%%%%%%%%

Turning to heavy-ion data, WA98~\cite{wa98-00} found a significant 
excess over $p$-$p$ ex\-trapolated primordial production in 
central $Pb$(158AGeV)+$Pb$ at SPS. The excess can be nice\-ly 
explained by thermal (QGP-) radiation with (average) initial temperatures 
$\bar T_0\ge 250$~MeV~\cite{Huo02}. However, if one allows for
a moderate Cronin effect in the primordial pQCD contribution, the
latter essentially exhausts the yield above $q_t\simeq 2$~GeV, 
and the QGP contribution becomes subdominant to the hadronic 
one~\cite{TRG04} (cf. right panel of Fig.~\ref{fig_phot}), a hierarchy 
quite similar to the NA50 dileptons (right panel of Fig.~\ref{fig_im}).
New data at low $q_t$ indicate significant 
excess over current calculations~\cite{TRG04}. The inclusion
of soft $\pi\pi$ Bremsstrahlung slightly improves the 
situation~\cite{TRG05} but additional effects seem to
be required, e.g. a softening of the ``$\sigma$''-meson or 
medium-modified $\Delta$ decays~\cite{Hees04b}.   

Preliminary data on direct photons in central $Au$-$Au$ at RHIC 
by PHENIX~\cite{Frantz04} show a substantial enhancement over
hard production at high $q_t$. The signal is not yet sensitive
to predicted thermal yields~\cite{Ras03,TRG04}.

%%%%%%%%%%%%%%%%%%%%%%%%%%%%%%%%%%%%%%%%%%%%%%%%%%%%%%%%%%%%%%%%%%%%%%%%
\section{Conclusions}
\label{sec_concl}
%%%%%%%%%%%%%%%%%%%%%%%%%%%%%%%%%%%%%%%%%%%%%%%%%%%%%%%%%%%%%%%%%%%%%%%%
Electromagnetic radiation from hot and dense QCD matter constitutes a 
valuable source of information on both its thermal and microscopic 
properties, in particular  (a) on early temperatures at masses/energies 
$\ge$~1.5~GeV, and (b) on hadronic in-medium effects, with a potential 
to study chiral restoration, at $M$,$q_0$~$\le$~1~GeV.  
Microscopically consistent calculations of the e.m. correlator
have been able to explain photon and dilepton spectra at the SPS in 
terms of thermal radiation with fair success.
Among the questions that have not been answered  to date are:
\begin{itemize}
%\item
%Can one establish relations between a deconfinement order parameter and 
%quarkonium observables?
\item
Is there a deeper reason for the agreement between in-medium hadronic
and (resummed) pQCD calculations for both dilepton and photon rates
around $T_c$? 
\item
How is chiral restoration realized in the vector-axialvector channel?
What is the role of baryons? Can one devise observables to 
distinguish different scenarios?
\item
If confirmed, what are possible explanations for the large enhancement
of low-energy photons observed by WA98?
\item
If confirmed by HADES, what is the origin of the thus far unexplained 
low-mass dilepton enhancement at the BEVALAC?
\end{itemize}
With the anticipated wealth of upcoming (precision) data over a wide 
range of energies (SIS, SPS, RHIC and LHC), the combination of theory
and phenomenology ought to provide answers.

\vspace{0.5cm}

\noindent{\bf Acknowledgment} \\
I thank the workshop organizers for the invitation 
to a very informative meeting. 
It is a pleasure to acknowledge my collaborators on the presented 
topics, C. Gale, L. Grandchamp, E. Greco, C.M. Ko, S. Turbide, 
H. van Hees and J. Wambach.

\end{document}